\documentclass{epl}
\usepackage{epsf}

\newcommand{\be}{\begin{equation}}

\newcommand{\ee}{\end{equation}}

\newcommand{\bea}{\begin{eqnarray}}

\newcommand{\eea}{\end{eqnarray}}

\title {Coherent vs incoherent pairing in
2D systems near  magnetic instability.}
\shorttitle{Coherent vs incoherent pairing\ldots}
\author{Ar. Abanov\inst{1}, Andrey V. Chubukov\inst{1}, \and 
A. M. Finkel'stein\inst{2}}
\shortauthor{Ar. Abanov \etal}
\institute{ 
\inst{1} Department of Physics, University of Wisconsin, Madison, WI 53706\\
\inst{2} Department of Codensed Matter Physics, Weizmann Institute of Science, 
Israel} 
\begin{document}
\maketitle  
\begin{abstract} 
We study the superconductivity in 
2D fermionic systems  near antiferromagnetic instability, 
 assuming that the pairing is mediated by spin fluctuations. 
This  pairing involves fully incoherent
fermions and diffusive spin excitations. We show that the  
competition between fermionic 
 incoherence and  strong pairing interaction yields the
  pairing instability temperature $T_{ins}$ 
 which increases and 
saturates as the magnetic correlation length $\xi \rightarrow \infty$. 
 We argue that in
this quantum-critical regime the pairing problem 
is qualitatively different from the BCS one.
\end{abstract} 
In this communication we analyse the 
pairing problem in 2D fermionic systems near antiferromagnetic instability.
 Our key goal is to investigate whether 
or not the closeness to antiferromagnetism is in conflict with the
 magnetically mediated $d-$wave pairing. This problem is rather peculiar as 
on one hand  the $d-$wave pairing amplitude increases
 at approaching the AFM instability due to softening of spin fluctuations
~\cite{scal}, while on the other hand, strong 
spin-mediated interaction destroys
 fermionic coherence~\cite{chubukov,ac} and therefore damages the ability of 
fermions to form Cooper pairs. 

We demonstrate that the competition between strong pairing interaction 
and the destruction
of fermionic coherence yields a pairing instability at a temperature 
$T_{ins}$ which increases and saturates when
 the magnetic correlation length $\xi \rightarrow \infty$. 
We  show that under certain conditions, 
 $T_{ins}$ is 
universal 
in the sense that it does not depend on the details of the 
electronic dispersion at energies comparable to 
 the fermionic bandwidth $W$, and 
 is determined by fermions located in a narrow region near  
hot spots - the points at 
the Fermi surface separated by the antiferromagnetic momentum ${\bf Q}$.
We assume in this paper that the Fermi surface does contain hot spots. 

We believe that the results of our analysis may be 
 applicable to both cuprates and heavy fermion materials.
For high $T_c$ 
cuprates, our results may be useful for
 understanding of the pseudogap physics in the underdoped regime, 
where the data show that the temperature
 when the system first displays
 superconducting precursors saturates at the lowest dopings~\cite{review}.
We conject that our $T_{ins}$ may be the onset of the pseudogap behavior, 
while the actual superconducting transition occurs
  at  a  smaller temperature.
 For heavy fermion materials,
our result may help understand the close correlation between 
the appearance of the superconductivity and an   
antiferromagnetic instability~\cite{lonzarich}.  

The point of departure for our analysis is 
the  spin-fermion model which
describes low-energy fermions interacting with
their collective spin degrees of freedom.  
This model can be viewed as the low-energy version of the lattice, 
Hubbard-type models, and is given by
\begin{equation}
{\cal H} =
  \sum_{{\bf k},\alpha} {\bf v_F} ({\bf k}-{\bf k}_F)  
 c^{\dagger}_{{\bf k},\alpha} c_{{\bf k},\alpha}
+ \sum_q \chi_0^{-1} ({\bf q}) {\bf S}_{\bf q} {\bf S}_{-{\bf q}} +
g \sum_{{\bf q,k},\alpha,\beta}~
c^{\dagger}_{{\bf k+ q}, \alpha}\,
{\bf \sigma}_{\alpha,\beta}\, c_{{\bf k},\beta} \cdot {\bf S}_{\bf -q}\, .
\label{intham}
\end{equation}
\begin{figure}[tbp]
\centerline{\epsfxsize=2.7in \epsfysize=0.6in
\epsffile{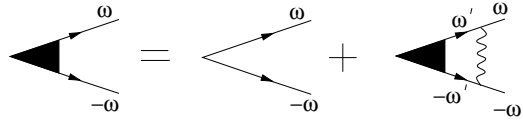}}
\caption{Diagrammatic representation for the pairing vertex. The solid and
wavy lines are fermionic and spin fluctuation propagators, respectively.}
\label{fig1}
\end{figure}
Here $c_{{\bf k}, \alpha} $ and ${\bf S}_{\bf q}$
describe fermions and collective bosonic spin
degrees of freedom, respectively, and  $g$  is the spin-fermion 
coupling constant. The three input parameters in the model are
 the Fermi velocity $v_F$, spin-fermion coupling $g$ 
(which near half-filling is of order Hubbard $U$), and  
the  spin correlation length $\xi$ defined via a
 bare  static spin susceptibility 
which is assumed to be peaked at the antiferromagnetic momentum ${\bf Q}$, 
i.e.,    
$\chi_0 ({\bf q}) = \chi_0 \xi^2/(1 + ({\bf q}-{\bf Q})^2 \xi^2)$.  
The dynamical part of the spin susceptibility
 comes from the interaction with the low-energy fermions 
and therefore is not an input.

The model of Eq. (\ref{intham}) yields a spin-mediated pairing interaction 
which singlet component is 
$\Gamma ({\bf q},\Omega) = 3 g^2 \chi ({\bf q},\Omega)$, 
where $\chi ({\bf q},\Omega)$ is 
the fully renormalized dynamical spin susceptibility. 
Near antiferromagnetic instability,
this interaction is attractive  in the $d_{x^2-y^2}$ channel~\cite{scal}.

A convenient way to study whether the spin-fermion
 interaction gives rise to a pairing at some $T_{ins}$ is to 
analyze a  linearized equation for the 
 fully renormalized $d-$wave pairing vertex $F$ with zero total momentum
and frequency.
 This vertex generally depends on relative  fermionic 
momentum $k$ and frequency $\omega$, i.e $F = F_k(\omega)$.
 In the ladder approximation which accuracy we discuss below,
the equation for $F_k (\omega_m)$ 
 takes the form (see Fig~\ref{fig1}).
\begin{equation}
F_k (\omega_m)=
F^{(0)}_k (\omega_m)
-T~\sum_{\omega ^{\prime_m }}~\int 
\frac{d^{2}k^{\prime }}{(2\pi )^{2}}F_{k^{\prime}} (\omega^{\prime}_m)~
G_{k^{\prime }}(\omega ^{\prime }_m)~G_{-k^{\prime }}(-\omega ^{\prime
}_m)~\Gamma (k-k^{\prime },\omega_m -\omega ^{\prime }_m).  \label{inteq}
\end{equation}
Here
$G_{k}(\omega )$ is the fully renormalized normal state
single-particle Green's function.
At $T = T_{ins}$, this equation should have a nontrivial solution even when
$F^{(0)}_k (\omega_m)=0$

To analyse Eq.(\ref{inteq}) we need to know the
fully renormalized  single-particle Green's function $G_k (\omega)$
and the pairing interaction $\Gamma (q,\Omega)$  
in the normal state. In 2D, the dimensionless 
coupling constant for 
Eq. (\ref{intham}) is 
$\lambda =3{\bar{g}}/(4\pi v_{F}\xi
^{-1})$, where $\bar{g}=g^{2}\chi _0$ is the effective spin-fermion interaction~\cite{chubukov}. Obviously, near a magnetic instability
 $\lambda \geq 1$, and a conventional perturbation 
expansion is inapplicable. It turns out, however, that 
one can resum perturbation series and obtain  
a  self-consistent solution for both 
$G_k (\omega)$ and $\Gamma (q,\Omega)$~\cite{chubukov,ac}.
This solution becomes exact in the formal limit $N \rightarrow \infty$ where 
 $N=8$ is the number of hot spots in the Brillouin zone. 
 Two of us have 
checked~\cite{ac2} that the corrections  to the spin-fermion 
vertex $g$ are small by $1/N$ and can be safely neglected.

The key  effect captured by the self-consistent solution
is the appearance of the small scale 
 $\omega_{sf}=9/(8 \pi N)~{\bar g}/\lambda^2 
\propto \xi^{-2}$~\cite{chubukov} which separates the 
regions of a Fermi liquid behavior at $\omega, T
 < \omega_{sf}$ and quantum-critical, non Fermi liquid 
behavior at $\omega, T > \omega_{sf}$.  Specifically, 
for electronic states near hot spots, $k \approx k_{hs}$,
\begin{eqnarray}
\chi ({\bf q},\Omega _{m})&=&\chi _{0}\xi ^{2}/
(1+({\bf q}-{\bf Q})^{2}\xi^{2}+
|\Omega _{m}|/\omega _{sf}) \nonumber \\ 
G_{k}^{-1}(\omega _{m})&=& i\omega_{m} Z_k (\omega_m)  
-\epsilon_{k}
\label{set}
\end{eqnarray} 
where
\begin{equation}
Z_k (\omega_m) = 1 + \frac{\pi T \lambda}{\omega_m}
~\sum_{n} \frac{\mbox{sign} \omega_n}{\sqrt{1 + \frac{|\omega_m -
 \omega_n|}{\omega_{sf}} + \left(\frac{\epsilon_{k+Q}}{v_F \xi^{-1}}\right)}}.
\label{z}
\end{equation}
Here $\epsilon_{k} = {\bf v}_{k} ({\bf k} - {\bf k}_{hs})$ and  
$|{\bf v}_F (k+Q)| = |{\bf v}_F (k)| = v_F$. 
At $T=0$ and $k = k_{hs}$, 
 $ Z (\omega _{m})= 1 + 2~\lambda/(1+\sqrt{1+|\omega _{m}|/\omega _{sf}})$.

Analyzing Eq. (\ref{set}) at $k=k_{hs}$, we find that at
 $\omega,T  \leq \omega_{sf}$,
$\chi ({\bf q},\Omega) \approx \chi_0 ({\bf q})$, and
 $G^{-1} (\omega)$ has a conventional Fermi liquid form
 $G^{-1} (\omega) \approx  \omega + i \mbox{sign} \omega (\omega^2 + 
\pi^2 T^2)/(4\omega_{sf})$. 
On the other hand, at $\omega, T > \omega_{sf}$, 
 \begin{eqnarray}
&&\chi^{-1} ({\bf q}, \Omega) \propto {\bar \omega} \left(\frac{q-Q}{q_0}\right)^2 - i \Omega \nonumber \\
&&G^{-1} \approx \omega + \left(i \pi T \lambda + 
(i|\omega|~{\bar \omega})^{1/2} f(T/|\omega|)\right) \mbox{sign} \omega
\label{g}
\end{eqnarray}
 where 
${\bar \omega} = 4 \lambda^2 \omega_{sf} = 9{\bar{g}}/(2\pi N)$, 
$q_0 = {\bar g}/(2\pi v_F)$,
 and $f(x)$ is a smooth function with $f(0)=1$ and $f(x \gg 1) \approx 
-1.52 \sqrt{ix}$. 
We see that spin fluctuations behave as
gapless diffusive modes and 
 fermionic excitations are fully incoherent.
This behavior is obviously a quantum-critical one.
Observe in this regard that ${\bar \omega}$ does not depend on 
the spin correlation length.
The scale ${\bar \omega}$ 
will play a central role in our further considerations.

The fermionic propagator also contains a
linear in $T$ term which does depend on $\xi$. This term, however,
comes from thermal spin fluctuations 
which contribute $n=m$ term to the frequency sum in Eq. (\ref{z}).
We will see that these fluctuations act as static impurities and
 do not affect $T_{ins}$. 

We first discuss in detail the pairing problem  when 
${\bar g}/v_{F}k_{F} \ll 1$, i.e., when $q_0 \ll k_F$.
We argue that in this case, 
the 
 pairing is dominated by fermions near hot spots and is insensitive to 
the system behavior at energies comparable to the bandwidth. Indeed,
 substituting the 
single particle Green's function, the spin 
susceptibility into Eq. (\ref{inteq}) and estimating  the momentum integral
 using a $d-$wave condition 
$F_k (\omega_m) = - F_{k+Q} (\omega_m)$,
 we find  that typical  $|Q-q|$ and $|k - k_{hs}|$ 
are of order $q_0$,
 i.e., are  much smaller than $k_F$. 

We also checked that 
for typical  momenta, $Z_k (\omega)$ and 
$F_k (\omega)$ are
weakly $k-$dependent and can be approximated by their values 
at a hot spot, $Z(\omega)$ and $F(\omega)$, respectively. 
Under these conditions, the momentum  integration can be performed exactly. 
The $N \rightarrow \infty$ limit is particularly
simple as  typical momenta
transverse to the Fermi surface  
are by a factor $1/N$ smaller than 
typical momenta along the Fermi surface. In this situation, the 
momentum integration is factorized: the one over transverse momenta
 affects only the fermionic Green's functions, while the 
integration over momenta along the Fermi surface affects
only the spin susceptibility. Performing the integration 
we obtain  
\begin{equation}
F(\omega_{m})= 
F^{(0)}(\omega _{m}) + 
\lambda \pi T \sum_{n}~\frac{F(\omega_n)}{|\omega_{n}| Z(\omega_{n})} 
~\frac{\sqrt{\omega_{sf}}}{\sqrt{\omega_{sf}+|\omega_{m}-\omega_{n}|}}
\label{new}
\end{equation}

Notice that the consequences of taking the $N \rightarrow \infty$ 
limit are the same as
of the Migdal theorem for phonon-mediated superconductors: 
one can (i) explicitly integrate over momentum in
the gap equation, and (ii)
neglect  corrections to $g$ and to ladder series. 
More precisely, 
 the $1/N$ smallness of the vertex corrections appears each time 
when these corrections involve
 fermions with momenta separated by $Q$~\cite{ac2}.
For the spin-fermion vertex, this is always the case, hence vertex corrections
 are small by $1/N$.  The
pairing vertex has a zero total momentum, and the ladder diagrams for this   
vertex, which give rise to Eq. (\ref{inteq}), 
do not contain $1/N$. However,
the corrections to ladder series from, e.g., crossed 
diagrams do involve fermions with momenta separated by $Q$, and
are small by $1/N$. 
From this perspective, our  analysis of the
spin-mediated pairing is quite similar to the Eliashberg analysis for
conventional superconductors~\cite{Eliash}. 

 We now analyse Eq. (\ref{new}). First we show that classical, thermal  
 spin fluctuations, which account for $i\pi T \lambda$ term in Eq. (\ref{g}), 
 do not affect $T_{ins}$. 
These fluctuations  account for the scattering with zero energy transfer
and therefore  act in the same way 
as impurities. 
 Accordingly, our argumentation
parallels the one which shows that nonmagnetic impurities do not 
affect $T_c$ in conventional superconductors~\cite{ag}.
Introducing ${\tilde F}_m  = F (\omega_m)/ \eta_m$ where 
$\eta_m = 1 + (\lambda \pi T/Z (\omega_m)/|\omega_m|)$,
we  explicitly rewrite Eq. (\ref{new}) as the equation for ${\tilde F}_m$
\begin{equation}
{\tilde F}_m=
F^{(0)}_m + 
 \lambda \pi T \sum_{n \neq m}~\frac{{\tilde F}_n}{|\omega_{n}| 
{\tilde Z}(\omega_{n})}
~\frac{\sqrt{\omega_{sf}}}{\sqrt{\omega_{sf}+|\omega_{m}-\omega_{n}|}}
\label{new1}
\end{equation}
where ${\tilde Z}$ is the same as in Eq. (\ref{z}) but without the 
contribution from
$m=n$ term in the frequency sum.
We see that Eq. (\ref{new1})
 contains only the contributions from 
quantum spin fluctuations.

We next discuss the form of the kernel in the r.h.s. of Eq. (\ref{new1}). 
We see that it contains two energy scales: $\omega_{sf} \propto \xi^{-2}$ 
and $\xi-$independent 
${\bar \omega} \gg \omega_{sf}$, which
is the upper cutoff for the $\sqrt{\omega}$ behavior of the fermionic 
propagator. 
For $|\omega| > {\bar \omega}$, the kernel converges
as $1/\omega^{3/2}$, i.e., 
the pairing problem does not extend above ${\bar \omega}$, 
which for ${\bar g} <  v_F k_F$ is still 
much smaller than the fermionic 
bandwidth.

The presence of the two 
 energies $\omega_{sf}$ and ${\bar \omega}$  
 raises the question on how $T_{ins}$ depends on $\xi$.  
To address this issue, consider the form of the kernel in Eq. (\ref{new1}) 
at different frequencies.
At $|\omega| < \omega_{sf}$, the system behaves as a Fermi 
liquid ($Z (\omega)  \approx 1 + \lambda$). 
In this frequency range,
the  kernel reduces to a constant, i.e., the pairing problem is of BCS 
type, with the 
 effective pairing coupling constant $\lambda/Z = \lambda/(1+\lambda)$ 
which never becomes large. If frequencies 
above $\omega_{sf}$ were not contributing to pairing, $T_{ins}$ would be
 of order 
$\omega_{sf} ~e^{-(1+\lambda )/\lambda }$, i.e., it would scale 
with $\omega_{sf}$. 
This is similar to what McMillan obtained for
conventional superconductors~\cite{mcmillan}. 

Consider next $|\omega| \geq \omega_{sf}$.  
Here the  pairing interaction (the 
last term in the r.h.s. of Eq. (\ref{new1}))
 becomes frequency dependent and gradually 
decreases compared to
its zero frequency value. At weak couplings, 
this decrease obviously
makes frequencies larger than $\omega_{sf}$ ineffective for pairing.
However,  at large $\lambda$ the situation  
is more tricky both in 
our case and for phonon superconductors~\cite{ad}.
The point is that for large $\lambda $, the mere
reduction of the pairing interaction above $\omega _{sf}$ is not sufficient -
one also has to neutralize the large
overall $\lambda $ factor in the r.h.s. of Eq.(\ref{new}). 
At $\omega <\omega _{sf}$, this overall $\lambda $ is neutralized 
by $Z (\omega_m) \approx  1+\lambda$.
However, above $\omega _{sf}$, $Z (\omega_m)$ decreases as 
$Z (\omega_m)\sim \lambda (\omega
_{sf}/|\omega_m|)^{1/2}$, and the effective coupling $\lambda/Z (\omega_m)$ 
increases.  
Simple power counting shows that this increase exactly balances the 
decrease of the pairing interaction such that  the
$1/|\omega|$ form of the pairing kernel survives  up to frequencies of 
order ${\bar \omega}$.
This may sweep the pairing instability to
a temperature $T_{ins} \sim {\bar \omega} \sim {\bar g}/N$.

To illustrate this point we introduce a dimensionless parameter  
$ n_T = ({\bar \omega}/(\pi T))^{1/2}$ and consider the
 limit $\omega_{sf} \rightarrow 0$. 
In this limit, Eq. (\ref{new1}) simplifies to 
\begin{equation}
{\tilde F}_{m}= 
F_{m}^{(0)}+
\frac{\alpha}{2} \sum_{n\neq m}\frac{{\tilde F}_{n}}{\sqrt{2|n-m|}
\sqrt{|2n+1|}}~\frac{n_T}{n_T+\sqrt{|2n+1|}}
\label{mats}
\end{equation}
 A fictitious parameter $\alpha $ ($=1$ in our case) is introduced  for
the subsequent perturbative analysis of this equation. 
We see that at low temperatures, i.e., large $n_T$, the kernel in 
Eq.(\ref{mats}) 
has a $1/n$ form typical for a pairing problem. 
On general grounds one
might expect that the pairing instability occurs at $n_T =O(1)$, i.e.,
at $T_{ins} \sim  {\bar \omega}$~\cite{phonons}. If this is the case, 
then  the pairing is dominated by frequencies where the fermionic excitations
display a fully incoherent quantum-critical behavior, i.e., the pairing 
is qualitatively different from that in a Fermi liquid.

The above argumentation is, however,
only suggestive as it is a'priori
unclear whether Eq. (\ref{mats}) has a nontrivial solution for any 
$n_T$. Indeed, on one hand, the  $1/\omega _{n}$ form of the kernel in 
Eq.(\ref{mats}) is typical for a pairing problem and  
gives rise to the logarithms in the ladder series. On the other hand, 
 this kernel 
depends not only on the running frequency as would be the case for BCS 
superconductivity, but also 
on the frequency transferred by the interaction. This last frequency
 serves as a lower cutoff for the logarithmical behavior.

 To get further insight into the problem 
 we assumed that $F^{(0)}_m$ is a constant and
analyzed  Eq.(\ref{mats}) for
various $\alpha$. 
We found that for small $\alpha $, when perturbative analysis
of the logarithmical series is valid, the  dependence of the kernel on the 
transverse frequency is crucial, and 
even at $T=0$, the summation of the series of logarithms
 give rise to a  power-law behavior 
${\tilde F}_{m}\propto F^{(0)}/|\omega _{m}|^{\alpha/2}$
rather than to a divergence. In other words, unlike BCS theory, 
at $\alpha \ll 1$, the 
logarithmical series do not give rise to a pairing instability. 

We find, however, that the convergence of the perturbation theory is
confined only to small $\alpha \ll 1$. Indeed, assume that at small
$\omega_m$, ${\tilde F}_{m}\propto |\omega _{m}|^{-1/4 + \beta}$.
Substituting this into  Eq.(\ref{mats}), we obtain an
 equation on $\beta$: $1 = (\alpha/2) \Phi (\beta)$, where
\begin{equation}
\Phi (\beta) = \frac{\pi^{3/2}}{\sqrt{2}}~\frac{1}{\Gamma(3/4 + \beta)
\Gamma(3/4 - \beta)}~\frac{1}{\cos{\pi \beta} - \cos{\pi/4}}.
\label{e1}
\end{equation}
For real $\beta$, $\Phi (\beta)$ is an even function of $\beta$, which
increases monotonically from $\Phi (0) 
 \approx 8.97$ and diverges at $\beta \rightarrow 1/4$ as
$\Phi (\beta) \approx 1/(1/4-\beta)$. 
For $\alpha \ll 1$, we find $\beta = 1/4 - \alpha/2$, i.e., 
${\tilde F}_{m}\propto |\omega _{m}|^{-\alpha/2}$, 
in agreement with the results of the 
summation of the logarithmical series.
As $\alpha$ increases, $\beta$ becomes smaller and reaches zero 
at $\alpha = \alpha_{cr} = 2\Phi^{-1}(0) \approx 0.22$. 
At larger $\alpha$, a solution with real $\beta$ is impossible, 
i.e., a perturbation theory breaks down. 
Instead, the condition $1 = (\alpha/2) \Phi (\beta)$ yields an 
imaginary $\beta = i \beta^*$ i.e
 ${\tilde F}_{m}\propto |\omega _{m}|^{-1/4} 
\cos{(\beta^*\log |\omega _{m}|)}$.
 Near $\alpha_{cr}$, we find $\beta^* \approx 1.2 (\alpha
-\alpha _{cr})^{1/2}$. The appearance of the oscillating 
solution at $T=0$ implies that 
the pairing susceptibility is {\it negative} 
for some $|\omega _{m}|$. 
This obviously signals that the normal state at 
$T=0$ is unstable against pairing. An  estimate of $T_{ins}$ may be
obtained  from a requirement that a temperature should exceed a maximum frequency where the  pairing susceptibility is  negative. 
For  sufficiently small $\beta^*$ this yields
$T_{ins} \propto {\bar \omega}~e^{- \pi/\beta^*}$.
We see therefore that for  $\alpha =1$, when $\beta^* =O(1)$, 
the attraction between fully
incoherent fermions is capable to produce a pairing instability at $%
T_{ins}\sim {\bar \omega} \sim {\bar{g}}/N$, 
as we conjected above, but this result has a non perturbative origin.
We also performed RG analysis of the leading $1/N$ vertex corrections and
found that they only slightly, by $O(1/N)$,
  change  $\alpha_{cr}$ which still remains
much smaller than 1.
 
\begin{figure}[tbp]
\centerline{\epsfxsize=2.7in \epsfysize=1.7in
\epsffile{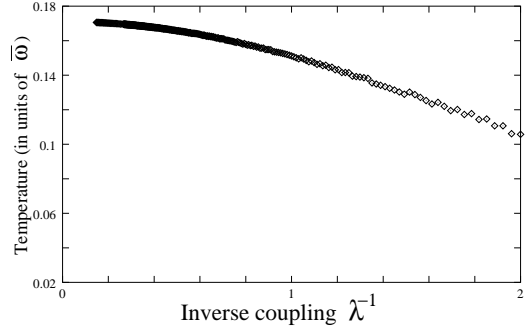}}
\caption{The results of the numerical solution of Eq~(\ref{new1}) for
different values of the coupling constant $\protect\lambda $.}
\label{fig3}
\end{figure}

To check this analysis, we solved our original Eq. (\ref{new1}) with $F^{(0)}_m =0$ numerically
for various $\lambda$. 
The results are presented in Fig. (\ref{fig3}). 
In the limit $\lambda \rightarrow \infty$ we found $T_{ins} \approx 0.17 
{\bar \omega}$.
It is interesting to observe that the weak dependence of 
$T_{ins}/{\bar \omega}$ 
on $\lambda$, which is an indicative of quantum critical 
superconductivity, persists down to $\lambda \sim 0.5$.  
This means that even at moderate  $\lambda$
the pairing instability  has a non-Fermi-liquid, quantum-critical origin. 

We now discuss the  momentum dependence of $F_k (\omega_m)$
at $T_{ins}$. This momentum dependence is likely to mimic that of a 
pairing gap
at $T < T_{ins}$~\cite{Pokr}. As we said above, $F 
(\omega_m)$ along the Fermi surface 
is weakly $k$ dependent at relative deviations from
 a hot spot by less than ${\bar g}/v_F k_F$
which is a small parameter in the theory. We checked that at larger 
deviations from a hot spot, 
 $F (\omega_m)$ rapidly decreases, 
as $1/(k -k_{hs})^2$. This means that 
for quantum-critical pairing, the $d-$wave pairing gap is more strongly 
confined to hot regions than a simple  $\cos k_x - \cos k_y$ form. 
This result is intuitively
obvious as  the very fact that the pairing problem is confined to hot 
spots implies
that the pairing state is a superposition of many
eigenfunctions from the $B_{1g}$ representation 
with almost equal partial amplitudes.
Simple manipulations with trigonometry show that in this situation, 
the slope of the gap near the nodes should be
smaller than the one inferred from the gap value at hot points assuming 
$\cos k_x - \cos k_y$ dependence of the gap. 
Notice, however, that this effect is non-critical, i.e., the width 
of the gap in $k-$space remains finite even if $\xi = \infty$.
  
Finally, we briefly discuss the 
situation at large spin-fermion interaction, when ${\bar g} \gg v_F k_F$, 
i.e., $ q_0 \gg k_F$ (see Eq. (\ref{g}). 
In this limit, the momentum integration extends over
 the whole fermionic bandwidth, and the presence of hot spots at the
Fermi surface becomes less relevant. 
The 
explicit evaluation of $T_{ins}$ is no longer possible, 
but  the reasoning along the same lines as above shows that $T_{ins}$
is independent on $\xi$ and scales as the largest typical frequency for 
the pairing problem. This typical frequency is obtained from the condition 
that maximum $|q-Q|$ are of order $k_F$, and is obviously
$J \sim (v_F k_F)^2/(N {\bar g})$.  
 
The analysis of the system behavior  below $T_{ins}$
 requires one to solve a
set of three coupled integral equations for the fermionic self-energy,
the anomalous vertex, and the spin susceptibility. Setting this aside for
a separate publication~\cite{future}, we merely 
argue here 
that the pairing state which emerges below $T_{ins}$ is highly unusual 
and has no analogs in BCS superconductors. Indeed, 
on one hand, $T_{ins}$ and hence the gap at $T=0$ are 
independent on $\xi$, 
on the other hand,  the resonance frequency 
of  the spin 
 mode 
scales as $\omega_{res} \sim  v_F\xi^{-1} \sim T_{ins}/\lambda$~\cite{ac},
 and  for $\lambda \gg 1$ is much smaller than the pairing gap.
In this situation, it is tempting to conject that superconducting
coherence may be destroyed by fluctuations not included in the Eliashberg treatment at $T_c < T_{ins}$, yielding a disordered 
region between $T_c$ and $T_{ins}$. This  
issue is, however, highly speculative and requires further study.  
 
We now briefly discuss the situation in cuprates. Near 
 half-filling, 
$v_F k_F$ scales with the fermionic bandwidth, while ${\bar g}$ is 
of order of the Hubbard $U$, hence $J$ and $T_{ins}$ (if 
${\bar g} \gg v_F k_F$) are	
 of order of the exchange integral of the
corresponding Heisenberg model.  
The actual situation in cuprates probably falls into an
intermediate regime ${\bar g} \geq v_F k_F$.
We emphasize however that for
$\omega_{sf} \sim 10-20 meV$, and $\lambda \sim 1$ extracted from NMR 
experiments at optimal doping~\cite{MP}, the universal result
(Fig. (\ref{fig3})) yields $T_{ins} \sim 10^2 -10^3 K$
which is a  reasonable estimate. 
The non-critical sharpening of the superconducting 
gap with underdoping is also consistent with the recent photoemission 
data~\cite{mesot}.
More detailed analysis requires a more precise knowledge
of both $\lambda$ and $\omega_{sf}$ for various doping concentrations.
 

Finally we discuss how our work is connected to earlier studies. 
The Eliashberg-type equations for magnetically mediated pairing have
 been analyzed several times in the literature~\cite{MP,MS,ML}, mostly
using the numerical technique. 
In particular, Monthoux and Lonzarich~\cite{ML} recently 
solved Eliashberg equations for large $\xi $ and for the Fermi surface 
with hot spots.
 They found that for large couplings, $T_{ins}$ 
likely 
 saturates at a finite value at $\xi =\infty $. This fully
agrees with our result for $T_{ins}$.
However, our key finding is the discovery that near antiferromagnetic 
instability,
the pairing problem is a quantum-critical one, and is 
qualitatively different from the BCS pairing. 
We also found that in the presence of hot spots at the Fermi surface, 
$T_{ins}$ 
is universal and does not
depend on the form of the pairing potential at lattice scales. 
 This physics
was not detected in earlier works~\cite{MP,MS,ML}.

It is our pleasure to thank G. Blumberg, L.P. Gor`kov, R.
Gatt, V. Kalatsky, D. Khveshenko, K. Kikoin, G. Kotliar,
Ph. Monthoux, M. Onellion, D. Pines, J. Schmalian, and A. Tsvelik for useful
discussions. The research was supported by NSFDMR-9629839 (Ar. A and A. Ch.),
 by The Israel Science Foundation - Center of Excellence Program  (A.M.
F.) and by Binational (US-Israel) Science Foundation.
A.M.F. is thankful to UW-Madison for the hospitality during the early stages
of this project.


\begin{thebibliography}{0}
\bibitem{phonons}
For phonon-mediated superconductivity, the pairing
interaction decreases inversely proportional to the square of the 
frequency, $Z (\omega_m) \propto \lambda \theta^2_D/|\omega_m|^2$,
where $\theta _{D}$ is the Debye frequency, 
and similar reasoning yields $T_{c}\propto \sqrt{\lambda }\theta _{D}$
 \protect\cite{ad}.
\end{thebibliography}
\end{document}